%% file: main.tex
\documentclass[%
reprint,
 twocolumn,
 amsmath,amssymb,
 aps,
 prl,
]{revtex4-2}

\usepackage[T1]{fontenc}
\usepackage{graphicx}
\usepackage{xcolor}
\usepackage[normalem]{ulem}
\usepackage{dcolumn}
\usepackage{bm}
\usepackage[mathlines]{lineno}
\usepackage{hyperref}
\usepackage{placeins}

\newcommand{\ppt}{p_{\rm T}}
\newcommand{\sqrtsNN}{$\sqrt{s_{_{\mathrm{NN}}}}$}
\def \auau  {Au+Au}

\begin{document}

\preprint{}

\title{Measurements of Proton High-Order Cumulants in \texorpdfstring{\sqrtsNN}~ = 3 GeV \auau{} Collisions and Implications for the QCD Critical Point }










\input{star-author-list-2021-09-15.iop}

\date{\today}

\begin{abstract}

We 
report cumulants of the proton multiplicity distribution from dedicated fixed-target \auau{} collisions at \sqrtsNN~$=3.0$ GeV, measured by the STAR experiment in the kinematic acceptance of rapidity ($y$) and transverse momentum ($\ppt$) within $-0.5 < y<0$ and $0.4 < \ppt <2.0 $ GeV/$c$. 
In the most central 0--5\% collisions, a proton cumulant ratio is measured to be $C_4/C_2=-0.85 \pm 0.09 ~(\rm stat.) \pm 0.82 ~(\rm syst.)$, which is 2$\sigma$ below the Poisson baseline with respect to both the statistical and systematic uncertainties. The hadronic transport UrQMD model 
reproduces our $C_4/C_2$ in the measured acceptance.
Compared to higher energy results and the transport model calculations, the suppression in $C_4/C_2$ is consistent with fluctuations driven by baryon number conservation 
and indicates an energy regime dominated by hadronic interactions. These data imply that the QCD critical region, if created in heavy-ion collisions, could only exist at energies higher than 3 GeV.
\end{abstract}

\maketitle

\input{paper_text}

We thank Dr. V. Koch, Dr. A. Sorensen and Dr. V. Vovchenko for interesting discussions. We thank the RHIC Operations Group and RCF at BNL, the NERSC Center at LBNL, and the Open Science Grid consortium for providing resources and support.  This work was supported in part by the Office of Nuclear Physics within the U.S. DOE Office of Science, the U.S. National Science Foundation, National Natural Science Foundation of China, Chinese Academy of Science, the Ministry of Science and Technology of China and the Chinese Ministry of Education, the Higher Education Sprout Project by Ministry of Education at NCKU, the National Research Foundation of Korea, Czech Science Foundation and Ministry of Education, Youth and Sports of the Czech Republic, Hungarian National Research, Development and Innovation Office, New National Excellency Programme of the Hungarian Ministry of Human Capacities, Department of Atomic Energy and Department of Science and Technology of the Government of India, the National Science Centre of Poland, the Ministry  of Science, Education and Sports of the Republic of Croatia and German Bundesministerium f\"ur Bildung, Wissenschaft, Forschung and Technologie (BMBF), Helmholtz Association, Ministry of Education, Culture, Sports, Science, and Technology (MEXT) and Japan Society for the Promotion of Science (JSPS).
\FloatBarrier

\bibliography{references}
\end{document}

%% file: star-author-list-2021-09-15.iop.tex
\author{
M.~S.~Abdallah$^{5}$,
B.~E.~Aboona$^{55}$,
J.~Adam$^{6}$,
L.~Adamczyk$^{2}$,
J.~R.~Adams$^{39}$,
J.~K.~Adkins$^{30}$,
G.~Agakishiev$^{28}$,
I.~Aggarwal$^{41}$,
M.~M.~Aggarwal$^{41}$,
Z.~Ahammed$^{61}$
% ...
}

\address{\rm{(STAR Collaboration)}}

% ...
\address{$^{2}$AGH University of Science and Technology, FPACS, Cracow 30-059, Poland}
% ...
\address{$^{5}$American University of Cairo, New Cairo 11835, New Cairo, Egypt}
\address{$^{6}$Brookhaven National Laboratory, Upton, New York 11973}
% ...
\address{$^{28}$Joint Institute for Nuclear Research, Dubna 141 980}
% ...
\address{$^{30}$University of Kentucky, Lexington, Kentucky 40506-0055}
% ...
\address{$^{39}$Ohio State University, Columbus, Ohio 43210}
% ...
\address{$^{41}$Panjab University, Chandigarh 160014, India}
% ...
\address{$^{55}$Texas A\&M University, College Station, Texas 77843}
% ...
\address{$^{61}$Variable Energy Cyclotron Centre, Kolkata 700064, India}
% ...

%% file: paper_text.tex
With the discovery of the quark-gluon plasma (QGP) at the Relativistic Heavy Ion Collider (RHIC)~\cite{BRAHMS:2004adc,PHOBOS:2004zne,PHENIX:2004vcz,STAR:2005gfr}, physicists are starting to investigate the phase structure of the QCD matter, especially in the high baryon density region.
The stark differences between the properties of QGP and lower energy nuclear matter draw interest to the thermodynamic processes, specifically those related to the nature of phase transitions~\cite{Stephanov:2011pb}. 
Experimenters can access the QCD phase diagram, expressed in temperature ($T$) and baryonic chemical potential ($\mu_{B}$), and search for phase boundaries by varying the heavy-ion collision energy.
At regions of equal baryon and anti-baryon density, $\mu_{B}$ = 0, theoretical approaches work well, with lattice QCD calculations predicting a smooth cross-over transition from hadronic matter to a QGP~\cite{Borsanyi:2013bia,Gupta:2011wh}. At finite $\mu_{B}$, where the baryon density is larger than the anti-baryon density, the existence and nature of the phase transition are not well understood.

The event-by-event fluctuations of conserved quantities such as net charge, net-baryon number, and net strangeness are predicted to depend on the non-equilibrium correlation length, $\xi$, and thus serve as indicators of critical behavior~\cite{PhysRevLett.91.102003}. Ideally, near the singular critical point, the correlation length could grow as large as the size of the system under study, provided sufficient time for the development.  
In heavy-ion collisions, however, effects from the finite size and limited lifetime of the hot nuclear system will limit the significance of signals~\cite{FRAGA}. 
A theoretical calculation suggests that $\xi$ may rise from $\sim\!0.5$ to $3$ fm in heavy-ion collisions, constrained by the size of the system~\cite{Berdnikov:1999ph}.
Experimentally, compared to other baryons, protons and anti-protons are measured with high efficiency~\cite{Aggarwal:2010wy} and have been shown to be reliable proxies for baryons and anti-baryons~\cite{PhysRevLett.91.102003}. 
Despite computational challenges at finite $\mu_B$~\cite{Bazavov:2017tot,Bellwied:2021nrt}, lattice QCD calculations have predicted a positive cumulant ratio of net-proton (proton minus anti-proton) $C_4/C_2$ for the formation of QGP matter at $\mu_B \leq$ 200 MeV.

Recent reports on net-proton fluctuation measurements from RHIC's first phase of the Beam Energy Scan program (BES-I)~\cite{Adam:2020unf,STAR:2021iop} have demonstrated the potential sensitivity of the cumulant ratios of $C_3/C_2$ and $C_4/C_2$ of the net-proton multiplicity distribution to the collision energy. Because of baryon number conservation, calculations from both hadron resonance gas models (HRG) of the canonical ensemble~\cite{Braun-Munzinger:2020jbk,Bass:1998ca,Vovchenko:2021kxx} and the Ultrarelativistic Quantum Molecular Dynamics (UrQMD)~\cite{Bass:1998ca,Bleicher:1999xi} transport model, which do not contain critical dynamics, produce a smooth energy dependence. Above a center of mass energy (\sqrtsNN) of 27 GeV, the Solenoidal Tracker at RHIC (STAR) collaboration's BES-I results agree well with these models~\cite{Adam:2020unf,STAR:2021iop}. However, in the energy range  7.7 < \sqrtsNN ~< 27 GeV from the top 5\% central Au+Au collisions at RHIC, STAR's results show a non-monotonic behavior as a function of \sqrtsNN~ with a significance of $3.1\sigma$~\cite{Adam:2020unf,STAR:2021iop}. Here, the centrality is a measure of the geometric overlap of two colliding nuclei and is defined by a charged particle multiplicity. At collision energies below \sqrtsNN~= 7.7 GeV, where net baryon densities are high, UrQMD predicts a suppression with respect to unity of $C_4/C_2$ for central events. For all energies, a gas of classical free particles (Poisson distribution) has a $C_4/C_2$ of 1. A remaining question is how the non-monotonic behavior continues in a higher baryon density region below \sqrtsNN~=7.7 GeV. 

In this paper, we report the cumulant ratios of proton multiplicity distributions in Au+Au collisions at \sqrtsNN~= 3.0 GeV. For the top 5\% central collisions, the dependence of cumulant ratios on the particle rapidity ($y$) and transverse momentum ($\ppt$) is presented along with comparisons to model calculations. At this energy, the anti-proton production is negligible ($\overline{p}/p \sim \exp({-2\mu_{B}/T_{\rm ch}})< 10^{-6}$)~\cite{Andronic:2017pug}, therefore only the proton multiplicity distribution is used in the analysis. 

The AGS-RHIC accelerator complex provided a gold beam with an energy of 3.85 GeV, incident on a gold target, corresponding to \sqrtsNN~= 3.0 GeV for Au+Au fixed-target collisions. 
At this energy, STAR's fixed-target mode (FXT)~\cite{STAR:2021beb,STAR:2020dav} covered the mid-rapidity for protons in the center-of-mass frame.   
The proton multiplicities are determined using the Time Projection Chamber (TPC) and Time of Flight detector (TOF) of the  STAR~\cite{Llope:2012zz}. 
The target was located 200.7 cm from the center of the TPC and of thickness 1.93 g/cm$^2$ ($0.25$ mm) corresponding to a 1\% interaction probability. 
The TPC measures both the trajectory and the energy loss ($dE/dx$) of a particle. The TPC is placed within a solenoidal magnetic field (0.5 T) and the particle momenta are calculated from their curvatures. 
For these data, RHIC was configured to circulate twelve bunches of $7 \times 10^9$ gold ions, which grazed the top of the gold target. 
To remove collisions between the beam and the beam pipe, event vertices are required to be less than 1.3 cm from the Au target along the beam line and less than 1.5 cm from the target radially from the mean collision vertex.
The analysis is performed with $1.4\times10^{8}$ events.

\begin{figure}[htbp]
    \centering
    \includegraphics[width=0.5\textwidth]{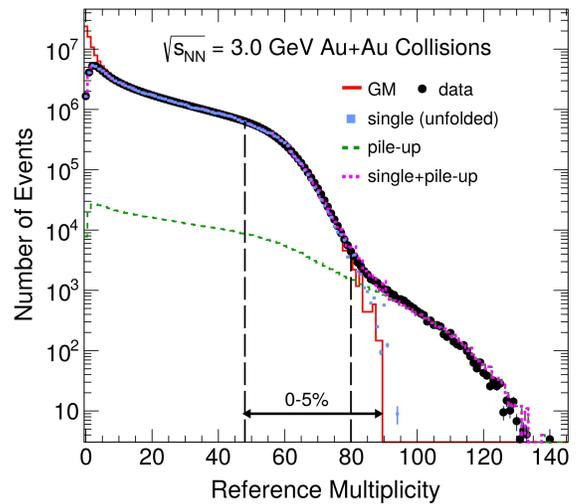}
    \caption{ Reference multiplicity distributions obtained from \sqrtsNN~= 3.0 GeV data (black markers), GM (red histogram), and single and pile-up contributions from unfolding. Vertical lines on markers represent statistical uncertainties. Single, pile-up and single+pile-up collisions are shown in solid blue markers, dashed green and dashed magenta curves, respectively. Analysis is performed on 0\%--5\% central events, indicated by a black arrow.}
    \label{fig:Pileup}
\end{figure}

The collisions are characterized by their centrality, inferred from reconstructed particle multiplicities (reference multiplicity). For this analysis, the reference multiplicity is the total number of tracks in the TPC uncorrected for efficiency loss, excluding baryons 
via $dE/dx$. The TPC covered all azimuthal angles and the pseudorapidity $\eta$ of $0<\eta<2$, in which $\eta\equiv -\ln[\tan(\theta/2)]$ and $\theta$ is the angle between the particle three-momentum and the beam axis in the lab frame. Proton tracks are excluded from the reference multiplicity to avoid self-correlations~\cite{Chatterjee:2019fey,Adam:2020unf, STAR:2021iop}. The reference multiplicity distribution shown in Fig.~\ref{fig:Pileup} is fit with a Monte Carlo Glauber model (GM) coupled with a two component particle production model~\cite{Miller:2007ri,STAR:2009sxc}. By integrating the GM fit, events are categorized into seven centrality classes: 0\%--5\%, 5\%--10\%, 10\%--20\%, ..., 50\%--60\%.
At reference multiplicities below $10$, the experimental data and the GM disagree due to inefficiency in the experimental trigger system. At multiplicities above $80$, double collision (pile-up) events dominate the multiplicity distribution. In addition to a pile-up correction discussed below, events above the reference multiplicity of 80 are removed from the 0\%--5\% centrality class. 

In the FXT collisions, due to finite target thickness, the pile-up is clearly present, see Fig.~\ref{fig:Pileup}. The cumulants are corrected for the effect of pile-up using an unfolding method~\cite{Nonaka:2020qcm,2021pileup_urqmd}. As a result, the single and double collisions are separated statistically.  
Figure ~\ref{fig:Pileup} shows the input GM fit (red curve) and the unfolded pile-up distribution (green dashed curve). The single collision distribution is extracted (blue points) from the measured distribution (black dots) and the unfolded pile-up distribution. The event-averaged pile-up probability, or total pile-up fraction, is determined to be ($0.46 \pm 0.09$)\% of all events and  ($2.10 \pm 0.40$)\% in the 0\%--5\% centrality class.

\begin{figure}[htbp]
    \centering
    \includegraphics[width=0.5\textwidth]{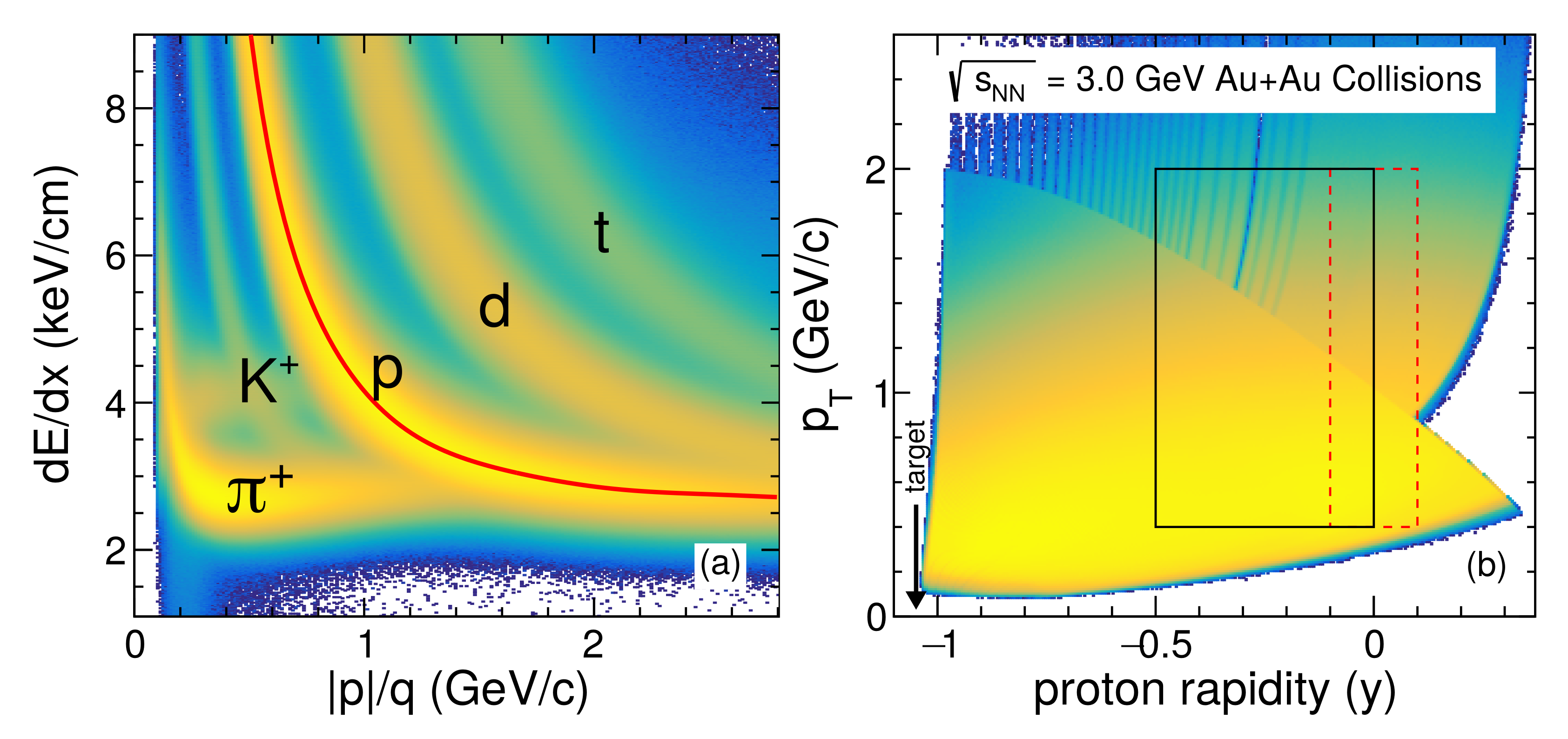}
    \caption{ Left panel (a):  $dE/dx$ versus particle rigidity measured in the TPC; pion, kaon, proton, deuteron, and triton bands are labeled. The theory prediction for protons is plotted in red. The electron peak is in between the pion and kaon bands.
    Right panel (b): Analysis acceptance in transverse momentum versus proton rapidity ($y$) in the center-of-mass frame of Au+Au collisions at \sqrtsNN~= 3.0 GeV. The black box indicates the acceptance within $-0.5 < y < 0$ and $0.4 < \ppt < 2.0$ GeV/$c$. The red dashed box indicates a narrower rapidity window $|y| < 0.1$, the largest possible symmetric rapidity window from this data set. In both panels, the yellow-to-blue color scale indicates the intensity. }
    \label{fig:PID}
\end{figure}

Figure~\ref{fig:PID}(a) shows $dE/dx$ versus the particle rigidity for all positively charged tracks in the STAR TPC. 
The pion, kaon, proton, and deuteron bands are labeled and a theoretical prediction~\cite{BICHSEL2006154} for the proton energy loss is shown in red.
Below rigidities of 2.0 GeV/$c$, the proton $dE/dx$ band is well separated and the TPC provides sufficient particle identification. To improve the particle identification for tracks with momenta above 2.0 GeV/$c$, TPC tracks are matched with TOF hits and a mass-squared cut of $0.6 < m^2 < 1.2$ (GeV/$c^2$)$^2$ is placed. The TOF requirement introduces a 60\% matching efficiency. The proton purity is required to be higher than 95\% at all rapidities and momenta for the subsequent cumulant analysis.

Figure~\ref{fig:PID}(b) displays the $\ppt$$-$$y$ acceptance in the center-of-mass frame for protons in fixed-target collisions at \sqrtsNN~= 3.0 GeV. The black box in Fig.~\ref{fig:PID}(b) indicates the acceptance window ($-0.5 < y < 0$, $0.4 < \ppt < 2.0$ GeV/$c$) used. The red dashed box shows the maximum symmetric rapidity window ($|y| < 0.1$) for the selected $\ppt$ region ($0.4 < \ppt < 2.0$ GeV/$c$). 
The target, depicted by a black arrow, is at rapidity $y = -1.05$. The diagonal discontinuity in Fig.~\ref{fig:PID}(b) is caused by the mass-squared cut above total momenta of 2.0 GeV/$c$ in the lab frame. The vertical line structure above 2.0 GeV/$c$, most prominent within $-1.0 < y < -0.2$, results from the geometry of the TOF modules. 

Experimentally measured proton multiplicity distributions are described by the central moments, $i.e.$, $\langle (\delta N)^2\rangle$, $\langle (\delta N)^3\rangle$ and so on. The symbol $\langle ... \rangle$ indicates the average over all the events, $N$ is the proton multiplicity in a given event, and $\delta N = N-\langle N \rangle$ is the deviation. The relations between the cumulants $C_n$ and the central moments are defined as:
\begin{equation}\label{eq:moment_to_cumulant}
\begin{aligned}
\rm{mean:} \  M            & =  \langle N \rangle                     &=&~~C_1, \\
\rm{variance:} \ \sigma^2  & = \langle (\delta N)^2 \rangle           &=&~~C_2, \\
\rm{skewness:} \ S         & =  \langle (\delta N)^3 \rangle/\sigma^3 &=&~~C_3/C_2^{3/2}, \\
\rm{kurtosis:}\ \kappa     & = \langle (\delta N)^4\rangle/\sigma^4-3        &=&~~C_4/C_2^2. 
\end{aligned}
\end{equation}

Ratios of the cumulants are often used to reduce volume dependence: $C_2/C_1 = \sigma^2/M$, $C_3/C_2 = S\sigma$, and $C_4/C_2 = \kappa\sigma^2$. An additional advantage is that the ratios of these cumulants can be readily compared with theoretical calculations of susceptibility \cite{EJIRI2006275,Friman:2011pf,KARSCH2011136,PhysRevLett.109.192302,PhysRevD.104.094047,PhysRevLett.111.062005,ALBA2014305} ratios $\sigma^2/M = \chi_2/\chi_1$, $S\sigma =\chi_3/\chi_2$, and $\kappa\sigma^2 = \chi_4/\chi_2 .$ 

The proton cumulants and ratios are corrected for detector inefficiency and background from pile-up collisions. 
The potential background from spallation in the beam pipe is reduced by the lower transverse momentum cut ($\ppt>0.4$ GeV/$c$). 
Detector efficiency corrections are performed on a ``track-by-track'' basis~\cite{PhysRevC.95.064912,Luo:2019}, where the proton reconstruction efficiency as a function of $\ppt$ and rapidity is applied as a weight to each track. The integrated proton track efficiency for the TPC detector is 95\% in the selected kinematic windows and centrality class (0\%--5\%).

All cumulant ratios are compared to the Poisson baseline for which cumulants of all orders are the same $C_n$$=$$M$ and the cumulant ratios are equal to one.
To suppress the spectator protons from entering the analysis, the maximum rapidity range is restricted to $-0.5 < y < 0$.
For the rapidity dependence measurement ($y_{\rm min} < y < 0$), the minimum rapidity ($y_{\rm min}$) is varied from $-0.5$ to $-0.2$ within $0.4<\ppt<2.0$ GeV/$c$.
For the transverse momentum dependence ($0.4<\ppt<\ppt^{\rm max}$), $\ppt^{\rm max}$ is varied from $0.8$ to $2.0$ GeV/$c$ within $-0.5 < y < 0$. 
The proton cumulants $C_1$ through $C_4$ are provided in the supplemental material \cite{SUPP}. 

The statistical uncertainties are obtained using a bootstrap approach~\cite{Luo:2013bmi, Pandav:2018bdx}. They are smaller than the marker size in the following figures. The systematic uncertainties are calculated from the uncertainty associated with the detector efficiency, the track selection criteria, and the pile-up correction.
To estimate the uncertainties in the track selection criteria, the mass-squared window, 
the number of TPC space points required, and the distance of closest approach (DCA) in 3-dimensions of the reconstructed track's trajectory to the primary vertex position was varied. The DCA was varied from 1--3 cm. The analysis used a DCA $<$ 3 cm cut.  
The uncertainty in the pile-up correction method is estimated by varying the pile-up fraction by its statistical uncertainty. 
For the top 5\% central collisions, the largest contributions to the systematic uncertainty for $C_4/C_2$ are from the pile-up correction ($\pm0.24$) and the DCA variation ($\pm0.78$). 

In a heavy-ion collision, the presence of noncritical fluctuations of the collision volume, \cite{Skokov:2012ds} also known as volume fluctuations (VF), may lead to an artificial enhancement in the measured cumulants \cite{Luo:2013bmi,He:2018mri}. As mentioned earlier, the information of collision centrality, expressed either in the fraction of total interaction cross section or in the averaged number of participating nucleons $\langle{\rm{N_{part}}}\rangle$,  is extracted  from the measured charged particle multiplicity distributions, see Fig.~\ref{fig:Pileup}. To achieve results  properly weighted by the event statistics, a centrality bin width correction (CBWC)~\cite{STAR:2021iop} is applied to all cumulants data discussed below. 
In comparison to BES-I, however, the centrality resolution in Au+Au collisions at \sqrtsNN~= 3.0 GeV is lower due to a decrease in the particle multiplicity. 
Therefore, volume fluctuation corrections (VFC)~\cite{Skokov:2012ds,BRAUNMUNZINGER2017114} are tested with both the hadronic transport model UrQMD~\cite{Bass:1998ca,Bleicher:1999xi} and Glauber model~\cite{Miller:2007ri}.

\begin{figure}[htb]
\centering
\includegraphics[width=0.4\textwidth]{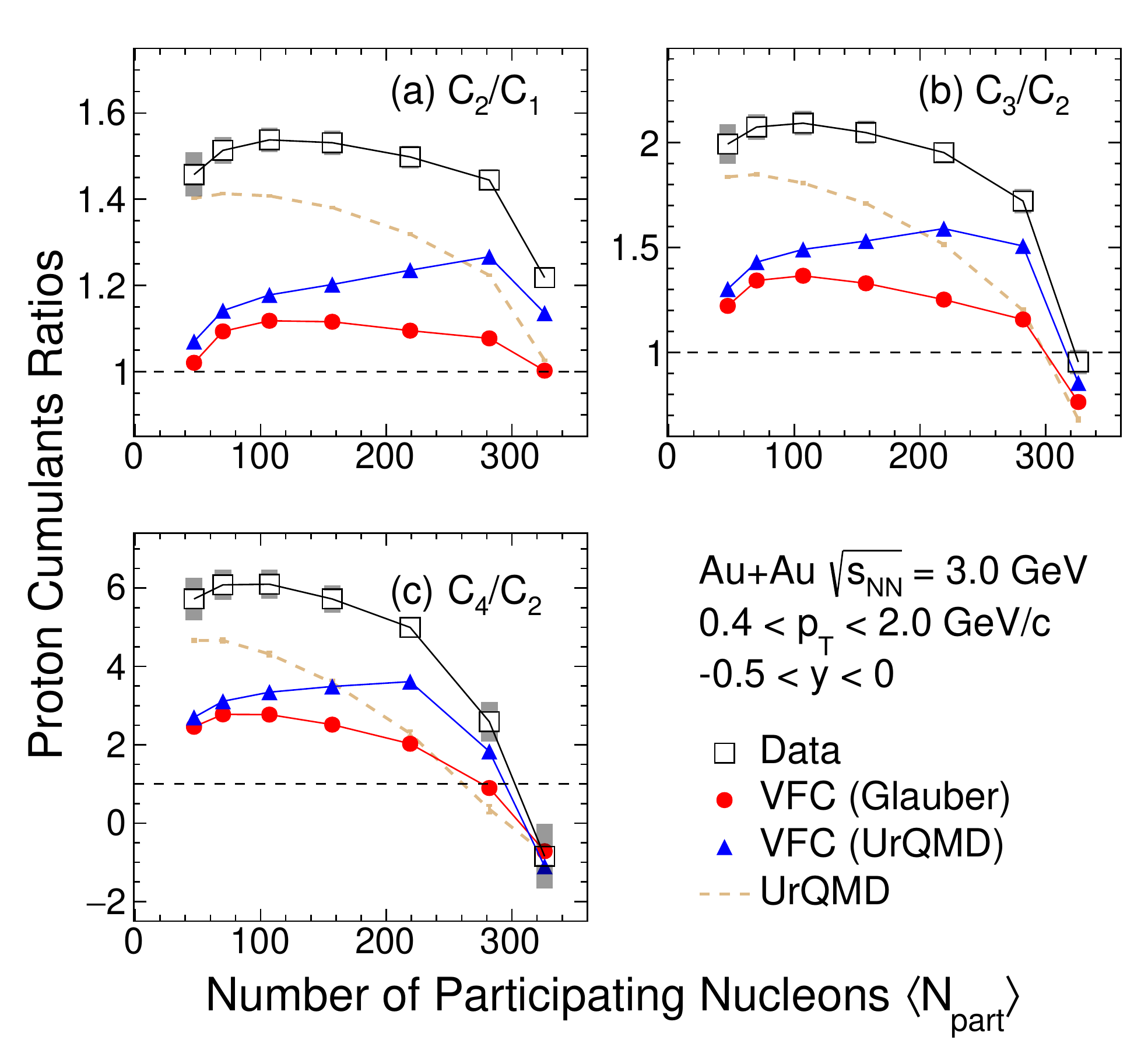}
\caption{Centrality dependence of the proton cumulant ratios for Au+Au collisions at \sqrtsNN~= 3.0 GeV. Protons are from $-0.5 < y < 0$ and $0.4 < \ppt < 2.0$ GeV/$c$. Systematic uncertainties are represented by gray bars. Statistical uncertainties are smaller than marker size. CBWC is applied to all cumulant ratios. While open squares represent the data without the VFC correction, blue triangles and red circles are the results with VFC using the $\langle N_{\rm{part}} \rangle$ distributions from the UrQMD and Glauber models, respectively. UrQMD model results are represented as gold dashed line.
} \label{fig:vfc_ratio}
\end{figure}

Figure~\ref{fig:vfc_ratio} depicts the cumulant ratios as a function of the average number of participating nucleons $\langle N_{\rm{part}}\rangle$. The data with VFC, using $ N_{\rm{part}}$ distributions extracted from UrQMD and Glauber models, and without VFC are shown as triangles, circles, and open squares, respectively. It is clear that the volume fluctuation correction shows a strong model dependence and affects the distribution, particularly in peripheral collisions. The respective dynamics in the UrQMD and Glauber model for charged hadron production lead to two different mappings from the measured final charged hadron multiplicity distributions to the initial geometry. This difference is likely the dominant source of the model dependence in the VFC. On the other hand, one can see in the figure that the difference between results with and without the VFC is small for higher order ratios $C_3/C_2$ and $C_4/C_2$ in the most central bin. As discussed in Refs.~\cite{Mackowiak-Pawlowska:2021sea,Xu:2016qzd}, the maximum number of participants, $N_{\rm{part}}^{\rm{max}}$ (394 for Au+Au collisions), suppresses the initial volume fluctuations. The trends in the centrality dependence of the cumulant ratios, $C_2/C_1, C_3/C_2$, and $C_4/C_2$, are well reproduced by the hadronic transport model UrQMD calculations, see gold dashed lines in Fig. 3.

\begin{figure}[htb]
\centering
\includegraphics[width=0.45\textwidth]{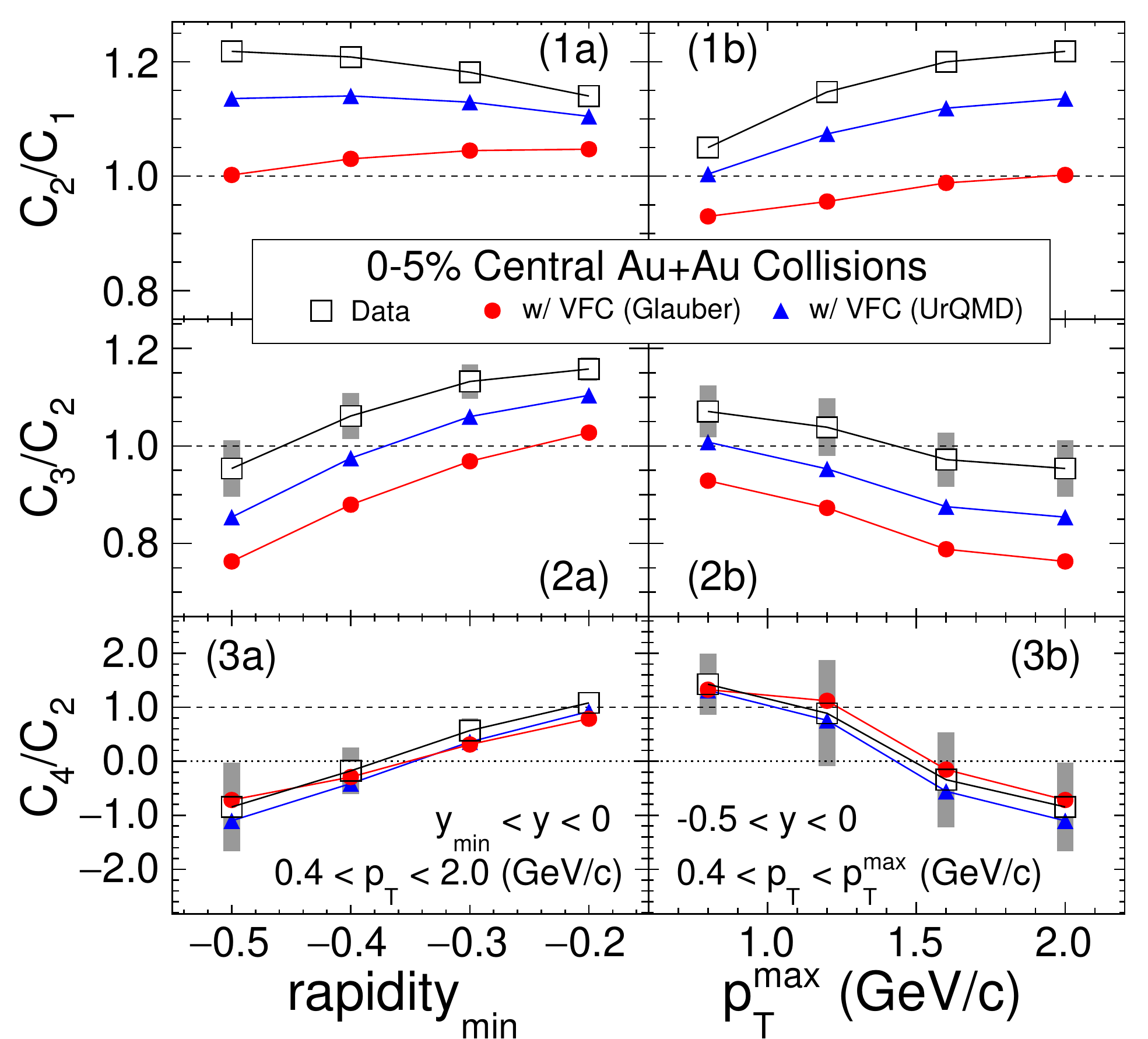}
\caption{Similar to Fig.~\ref{fig:vfc_ratio}: Rapidity and transverse momentum dependence of the proton cumulant ratios for 0\%--5\% central collisions. Black-squares, red-dots and blue-triangles stand for data without and with the VFC using Glauber and UrQMD, respectively.    
} \label{fig:Cn}
\end{figure}

Figure~\ref{fig:Cn} depicts the cumulant ratios as a function of rapidity $y$ and transverse momentum $\ppt$ in 0\%--5\% central collisions without and with the VFC.
It is expected~\cite{Ling:2015yau,Bzdak:2017ltv,Brewer:2018abr} that the cumulant ratios approach the Poisson baseline in the limit of small acceptance.
For $C_3/C_2$, the ratios with the VFC (UrQMD) and without the VFC deviate from the Poisson baseline at the narrow rapidity windows. The VFC (Glauber) ratio approaches unity as the acceptance is decreased. For the $C_4/C_2$ ratio, the VFC has a negligible effect in the most central bin.
Therefore, $C_4/C_2$ is reported without VFC in the discussions below. In the central 0\%--5\% collisions, as shown in Fig.~\ref{fig:Cn}, one obtains $C_4/C_2 = -0.85 \pm 0.09~(\rm stat.) \pm 0.82~(\rm syst.)$ in the kinematic acceptance of $-0.5 < y< 0$ and $0.4 <\ppt < 2.0$ GeV/$c$. The UrQMD model qualitatively reproduces the acceptance dependence of the data, see Fig.~6 in the supplemental material~\cite{SUPP}. 

\begin{figure}[htb]
\centering
\includegraphics[width=0.5\textwidth]{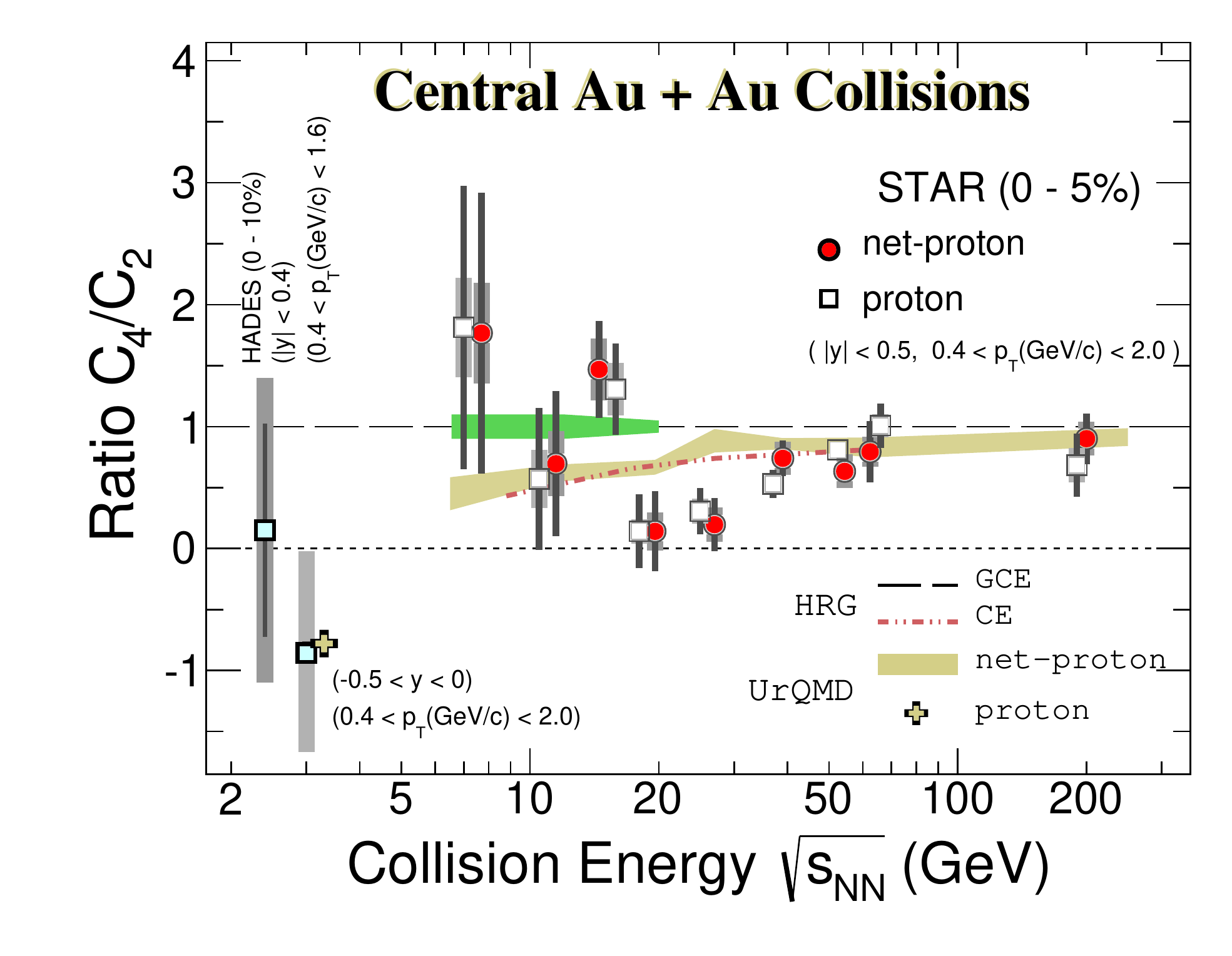}
\caption{Collision energy dependence of the ratios of cumulants, $C_4/C_2$, for proton (squares) and net-proton (red circles) from top 0\%--5\% Au+Au collisions at RHIC~\cite{Adam:2020unf,STAR:2021iop}. The points for protons are shifted horizontally for clarity. The new result for proton from \sqrtsNN~= 3.0 GeV collisions is shown as a filled square. HADES data of \sqrtsNN~= 2.4 GeV 0\%--10\% collisions~\cite{Adamczewski-Musch:2020slf} is also shown. The vertical black and gray bars are the statistical and systematic uncertainties, respectively. In addition, results from the HRG model, based on both Canonical Ensemble (CE) and Grand-Canonical Ensemble (GCE), and transport model UrQMD are presented.  }\label{fig:vsenergy}
\end{figure}

A non-monotonic energy dependence of the net-proton $C_4/C_2$ was reported for 0\%--5\% central Au+Au collisions at \sqrtsNN~= 7.7--200 GeV~\cite{Adam:2020unf, STAR:2021iop}. A similar energy dependence of the $C_4/C_2$ of protons is also evident (open squares in Fig.~\ref{fig:vsenergy}). Though a minimum appears around 20 GeV, both the $C_4/C_2$ ratio of protons and net-protons at 7.7 GeV are close to unity, albeit the large statistical uncertainties. Meanwhile, the $C_4/C_2$ value for Au+Au collisions at \sqrtsNN~= 3.0 GeV is around $-1$. The negative value of the proton $C_4/C_2$ is reasonably reproduced by the transport model UrQMD~\cite{Bass:1998ca, Bleicher:1999xi}. The HADES result of Au+Au at \sqrtsNN~$=2.4$ GeV in top 10\% central collisions~\cite{Adamczewski-Musch:2020slf} is shown in the figure as filled square. Overall, our ratio of $C_4/C_2$ (also $C_2/C_1$ and $C_3/C_2$) is consistent with the HADES data within uncertainties although detailed comparison should be done within same acceptance. It is worthy to note that we do not observe the large variations in the rapidity width as reported by HADES~\cite{Adamczewski-Musch:2020slf}.

The study of cumulant ratios in heavy-ion collisions has motivated several QCD inspired model calculations~\cite{Stephanov:2011pb}, which report a similar 
oscillation pattern around the critical point due to the symmetry 
of the medium~\cite{Schaefer:2011ex,Chen:2015dra,Fu:2016tey,Shao:2017yzv,Li:2018ygx,Mroczek:2020rpm}.  However, due to the stochastic nature of heavy-ion collisions, the finite lifetime and size of the system \cite{Wang_2011}, and dynamical effects such as the critical slowing will smear the ``critical point'' to a region in collision energy~\cite{Weil:2016zrk,Shen:2020mgh}. 

Poisson statistics and the Grand Canonical Ensemble (GCE) model predict that $C_4/C_2$ is 1.
Because of baryon number conservation, calculations from models without critical dynamics such as the Canonical Ensemble (CE)~\cite{Braun-Munzinger:2020jbk} and UrQMD~\cite{Bass:1998ca,Bleicher:1999xi} show a characteristic suppression with respect to the Poisson baseline in the net-proton $C_4/C_2$ when the collision energy is decreased, as seen in Fig.~\ref{fig:vsenergy}. The same experimental cuts on event centrality, rapidity, and transverse momentum have been applied to these calculations. It is worth noting that if the rapidity window is extended to $|y|<0.5$, the UrQMD model predicts a value of $C_4/C_2 \approx -4$ for proton in central Au+Au collisions at \sqrtsNN~= 3.0 GeV.  
Compared to results from higher energy collisions, the suppression of the $C_4/C_2$ ratio in central Au+Au collisions at 3.0 GeV is stronger due to baryon stopping and conservation.
Recently, a hadronic equation of state for 3.0 GeV Au+Au collisions was shown to be applicable, using the measurement of collective flow parameters~\cite{shaoweilan}.
While the low  $C_4/C_2$ value observed at the energy can be explained by fluctuations driven by baryon number conservation in a region of high baryon density where hadronic interactions are dominant, the non-monotonic variation~\cite{Adam:2020unf,Sorensen:2020ygf,STAR:2021iop} observed at higher collision energies is not demonstrated by the dynamics in non-critical models such as UrQMD. Precision data from the energy window of 3 $<$~\sqrtsNN~$<$ 20 GeV are needed in order to explore the possibility of critical phenomena.

In summary, cumulant ratios of proton multiplicity distribution from \sqrtsNN~= 3.0 GeV Au+Au collisions are reported. The new data are measured by the STAR experiment configured in fixed-target mode. At this collision energy, large effects due to the initial volume fluctuation are observed in the cumulant ratios except in the most central 0\%--5\% bin. The protons are measured with the acceptance $-0.5 < y < 0$ and $0.4 < \ppt < 2.0$ GeV/$c$. The rapidity and transverse momentum dependencies of the cumulant ratios $C_2/C_1$, $C_3/C_2$, and $C_4/C_2$ are presented. A suppression with respect to the Poisson baseline is observed in proton $C_4/C_2 = -0.85 \pm 0.09~(\rm stat) \pm 0.82~(\rm syst) $
in the most central 0\%--5\% collisions at 3 GeV and the UrQMD model reproduces the observed trend in the centrality dependence of the cumulant ratios including $C_2/C_1, C_3/C_2$, and $C_4/C_2$. 
This new result is consistent with fluctuations driven by baryon number conservation at the high baryon density region.